\documentclass[journal]{IEEEtran}

\ifCLASSINFOpdf
\else
   \usepackage[dvips]{graphicx}
\fi
\usepackage{url}
\usepackage{graphicx}
\usepackage{bm}
\usepackage{algorithmic}
\usepackage{algorithm}
\usepackage{array}
\usepackage[caption=false,font=normalsize,labelfont=sf,textfont=sf]{subfig}
\usepackage{textcomp}
\usepackage{stfloats}
\usepackage{url}
\usepackage{verbatim}
\usepackage{graphicx}
\usepackage{cite}
\usepackage{cite}
\usepackage{amsmath,amssymb,amsfonts}
\usepackage{algorithmic}
\usepackage{graphicx}
\usepackage{bm}
\usepackage{textcomp}
\usepackage{xcolor}
\usepackage{float}
\usepackage{multirow}
\usepackage{colortbl}
\usepackage{xcolor}

\usepackage{graphicx}

\begin{document}

\title{P-norm based Fractional-Order Robust Subband Adaptive Filtering Algorithm for Impulsive Noise and Noisy Input}

\author{Jianhong Ye, Haiquan Zhao, \emph{Senior} \emph{Member}, \emph{IEEE}, Yi Peng
	\thanks{This work was supported by the National Natural Science Foundation
		of China (NSF) under Grant 62171388, Grant 61871461, and Grant
		61571374. (Corresponding author: Haiquan Zhao).
		
		The authors are with the Key Laboratory of Magnetic Suspension Technology and Maglev Vehicle, Ministry of Education, School of Electrical Engineering, Southwest Jiaotong University, Chengdu, China. E-mail address: yjh\_zcl@163.com (J. Ye), hqzhao\_swjtu@126.com (H. Zhao), pengyi1007@163.com (Y. Peng).}}
\maketitle

\begin{abstract}
Building upon the mean $p$-power error (MPE) criterion, the normalized subband $p$-norm (NSPN) algorithm demonstrates superior robustness in $\alpha$-stable noise environments ($1<\alpha \leq 2$) through effective utilization of low-order moment hidden in robust loss functions. Nevertheless, its performance degrades significantly when processing noise input or additive noise characterized by $\alpha$-stable processes ($0<\alpha \leq 1$). To overcome these limitations, we propose a novel fractional-order NSPN (FoNSPN) algorithm that incorporates the fractional-order stochastic gradient descent (FoSGD) method into the MPE framework. Additionally, this paper also analyzes the convergence range of its step-size, the theoretical domain of values for the fractional-order $\beta$, and establishes the theoretical steady-state mean square deviation (MSD) model. Simulations conducted in diverse impulsive noise environments confirm the superiority of the proposed FoNSPN algorithm against existing state-of-the-art algorithms.
\end{abstract}


\IEEEpeerreviewmaketitle

\section{Introduction}
As a prevalent optimization technique, the stochastic gradient descent method has been extensively employed in adaptive filtering algorithm design. The normalized least mean square (NLMS) algorithm \cite{slock1993convergence}, operating under the minimum mean square error criterion, finds extensive applications in active noise control \cite{chen2022distributed, shah2014fractional, kranthi2024charbonnier, padhi2017design, chang2014feedforward}, dereverberation \cite{huang2008dereverberation}, and acoustic echo cancellation \cite{morgan1995slow, paleologu2015overview} scenarios owing to its structural simplicity and computational efficiency. To address the inherent limitation of slow convergence under strongly correlated input signals, Lee and Gan extended the NLMS algorithm to the subband domain \cite{lee2004improving}. The resultant normalized subband adaptive filter (NSAF) algorithm employs analysis filter banks to decompose the correlated input signals into approximately white subband signals. This advancement achieves accelerated convergence rates compared to conventional NLMS implementations while maintaining comparable computational complexity. However, the performance of both the NSAF and NLMS algorithms degrades significantly under impulsive noise environments. Generally, such noises can be effectively modeled using a symmetric $\alpha$-stable random process whose characteristic function is expressed as: \cite{bergamasco2012active}:
\begin{align}
	\footnotesize
	\label{000} 
	\phi(t)=\text{exp}\big(-\zeta \lvert t\lvert^{\alpha}\big).
\end{align}
In \eqref{000}, $\zeta>0$ is the scale parameter, and the characteristic exponent $\alpha \in(0,2]$ controls the impulsiveness of the noise. Many robustness strategies have been proposed to handle such impulsive noises, such as M-estimate \cite{huang2022propor}, mixed-norm \cite{zayyani2024robust}, information-theoretic learning \cite{chen2021asymmetric, kumar2022gen}, sign algorithm \cite{zheng2019steady}, and so on. To enhance the robustness of the NSAF algorithm under impulsive noise environments, the maximum correntropy criterion-based NSAF (MCC-NSAF) algorithm has been developed through information-theoretic learning principles, demonstrating significant robustness against impulsive disturbances \cite{10985768, zhao2023robust}. Another approach, the normalized subband $p$-norm (NSPN) algorithm \cite{ye2024p}, achieves some desirable results in $\alpha$-stable noise environments; however, it demonstrates degraded robustness when $\alpha$ belongs to $0<\alpha\leq1$. 

Under more extreme noise conditions characterized by $\alpha$-stable distributions in both input and noise signals, the algorithms mentioned above demonstrate limited effectiveness in achieving satisfactory estimation accuracy. Fractional-order filters have demonstrated unique advantages in addressing this challenge \cite{li2022performance}, \cite{xie2021enhanced}, \cite{yang2022fractional}, \cite{kar2025fractional}, \cite{gogineni2020fractional, zhang2021optimal, wang2025fractional}. By integrating fractional-order calculus principles with the fractional-order stochastic gradient descent (FoSGD) framework \cite{jumarie2013derivative}, these filters exhibit superior convergence rates and reduced steady-state errors in dealing with such complex noise environments. For instance, the fractional-order normalized least mean $p$-norm (FoNLMP) algorithm achieves stable convergence when the input and noise following $\alpha$-stable distribution \cite{luo2021fractional}. Furthermore, the FoSGD approach was extended to the maximum Versoria criterion (MVC), thereby generating the FoMVC algorithm \cite{abdelrhman2024fractional}. Recently, to further improve the convergence performance of the FoMVC algorithm, the authors proposed fractional-order generalized Cauchy kernel loss (FoGCKL) and enhanced batch FoGCKL (EB-FoGCKL) algorithms \cite{cui2025enhanced}.

To address the limitations of the NSPN algorithm in $0<\alpha\leq1$ regime and improve its robustness against $\alpha$-stable process inputs, this paper proposes a fractional-order NSPN (FoNSPN) algorithm. Our key contributions include:
\begin{itemize}
	\item A novel robust fractional-order NSPN (FoNSPN) algorithm is developed to address the challenges that arise when dealing with highly correlated inputs and when the input and noise signals are modeled by the $\alpha$-stable processes.
	\item The mean convergence stability conditions and steady-state mean square deviation model of the FoNSPN algorithm have been specified.
\end{itemize}

\section{Overview of NSPN algorithm}
\label{sec:guidelines}

Within the system identification framework, the input-output mapping of the unknown system $\bm{h}_0$ is governed by the following relationship:
\begin{align}
	\label{001} 
	d_k = \bm{x}_k^{\text T}\bm{h}_0 + n_k,
\end{align}
where $d_k$ denotes the desired signal, $\bm{x}_k$ represents the input vector with a size of $L\times 1$, $n_k$ is the additive noise with variance $\sigma_{n_k}^2$, and $(.)^{\text T}$ is the transpose operator. 

In the subband structure, by filtering the signals {$\bm{x}_k$, $d_k$, and $n_k$} through the analysis filters whose impulse responses are denoted as $\{\bm{f}_i\}_{i=1}^N$ with a size of $D\times 1$, where $N$ represents the number of the subband, we obtain the subband input vector $\bm{x}_{i,k}$, the subband desired signal $d_{i,k}$, and subband noise signal $n_{i,k}$, respectively. The subband error signal $e_{i,k}$ can be calculated by \cite{lee2004improving}
\begin{equation}
	\begin{split}
		\begin{array}{rcl}
			\begin{aligned}
				\label{002}
				e_{i,k} =d_{i,k} - \bm{x}_{i,k}^{\text T}\hat{\bm h}_k&=\bm{x}_{i,k}^{\text T}\big(\bm{h}_0-\hat{\bm h}_k\big)+n_{i,k}\\&=\bm{x}_{i,k}^{\text T}\widetilde{\bm{h}}_k+n_{i,k},
			\end{aligned}
		\end{array}
	\end{split}
\end{equation}
where $\hat{\bm h}_k$ is the estimate of $\bm{h}_0$, $\widetilde{\bm{h}}_k\overset{\bigtriangleup}{=}\bm{h}_0-\hat{\bm h}_k$ denotes the weight deviation vector. Furthermore, by updating the adaptive filter $\hat{\bm h}_k$ through the signals $\bm{x}_{i,k}$ and $e_{i,k}$ in the subband domain, we thereby obtain the traditional NSAF algorithm \footnote{The weight update equation is $\hat{\bm h}_{k+1}=\hat{\bm h}_k+\mu\sum_{i=1}^N\frac{\bm{x}_{i,k}e_{i,k}}{\lvert\lvert \bm{x}_{i,k}\lvert\lvert_2^2}$.} \cite{lee2004improving}. Clearly, when additive noise $n_k$ is modeled as impulsive noise, the outliers contained in the subband error signal $e_{i,k}$ are introduced into the update process of the NSAF algorithm, thereby preventing stable convergence. To address this issue, our previous work proposed the NSPN algorithm \cite{ye2024p} and formulated a robust loss function inspired by the MPE criterion as
\begin{equation}
	\begin{split}
		\begin{array}{rcl}
			\begin{aligned}
				\label{003}
				J(k)=\sum_{i=1}^N \lvert e_{i,k}\lvert^p,
			\end{aligned}
		\end{array}
	\end{split}
\end{equation}
where $1\leq p<\alpha\leq2$. Using the stochastic gradient descent method, the NSPN algorithm can be formulated as
\begin{equation}
	\begin{split}
		\begin{array}{rcl}
			\begin{aligned}
				\label{004}
				\hat{\bm h}_{k+1}=\hat{\bm h}_k+\mu\sum_{i=1}^N\frac{\bm{x}_{i,k}\lvert e_{i,k}\lvert^{p-1} \text{sgn}(e_{i,k})}{\lvert\lvert \bm{x}_{i,k}\lvert\lvert_p^p},
			\end{aligned}
		\end{array}
	\end{split}
\end{equation}
where $\text{sgn}(.)$ denotes the sign function and $\mu$ is the step-size.

\textbf{Remark 1}. When $p\geq1$, the cost function $J(k)$ described in \eqref{003} is a convex function with a unique minimum point, which guarantees stable convergence of the NSPN algorithm in impulsive noise environments ($\alpha\in(1,2]$). Conversely, the cost function $J(k)$ is not first-order differentiable, and the stochastic gradient descent method cannot be used to minimize $J(k)$ in the scenarios of $0\leq p<\alpha\leq1$, resulting in significant performance degradation under such noise environments. Furthermore, the presence of impulsive input signals introduces outliers that propagate through the weight update process via $\bm{x}_{i,k}$, adversely affecting algorithmic stability through erroneous gradient updates. 

\section{Proposed Algorithm}
To solve the limitations of the NSPN algorithm, we employ the FoSGD method to formulate the $\beta$-order gradient model for the weight vector as:
\begin{equation}
	\begin{split}
		\begin{array}{rcl}
			\begin{aligned}
				\label{005}
				\hat{\bm h}_{k+1}=\hat{\bm h}_k-\mu \nabla^{\beta}J(k),
			\end{aligned}
		\end{array}
	\end{split}
\end{equation}
where $\nabla^{\beta}(.)$ denotes $\beta$-order gradient operator.
\subsection{Fractional-Order Calculus}
To facilitate the subsequent derivation, we have introduced the following lemmas. 

\textbf{Lemma 1:} \textit{The fractional derivative of any compound function } $M\big(y(x)\big)$ \textit{follows the chain rule defined as \cite{jumarie2013derivative}}
\begin{equation}
	\begin{split}
								\small
		\begin{array}{rcl}
			\begin{aligned}
				\label{006}
				\Big[{M}\big(y(x)\big)\Big]^{(\beta)}=M_y^{(\beta)}(y)(y_x^{'})^{\beta},
			\end{aligned}
		\end{array}
	\end{split}
\end{equation}
where $\beta$ represents the order of the fractional derivative, and $y_x^{'}$ denotes the first-order derivative with respect to $x$.

\textbf{Lemma 2:} \textit{ For the fractional derivative of any exponential function, the following equality holds: \cite{jumarie2013derivative}} 
\begin{equation}
	\begin{split}
		\begin{array}{rcl}
			\begin{aligned}
				\label{007}
				D^{\beta}x^n=\Gamma(n+1)\Gamma^{-1}(n+1-\beta)x^{n-\beta},\;\; n>0,
			\end{aligned}
		\end{array}
	\end{split}
\end{equation}
where $\beta<n$, $\Gamma(.)$ is the Gamma function, and $D^{\beta}(.)$ represents the fractional differential operator.
\subsection{Fractional-Order NSPN Algorithm}
By utilizing the fractional-order chain rule defined in \eqref{006}, the term $\nabla^{\beta}J(k)$ in \eqref{005} can be calculated by
\begin{equation}
	\begin{split}
				\small
		\begin{array}{rcl}
			\begin{aligned}
				\label{008}
				\nabla^{\beta}J(k)=D^{\beta}J(k)\Bigg[\frac{\partial\big(\lvert e_{i,k}\lvert\big)}{\partial \hat{\bm h}_k}\Bigg]^{\beta}.
			\end{aligned}
		\end{array}
	\end{split}
\end{equation}
Then, based on \textbf{Lemma 2}, the first part in the right-hand side of \eqref{008} can be obtained as
	\begin{equation}
		\begin{split}
								\small
			\begin{array}{rcl}
				\begin{aligned}
					\label{009}
					D^{\beta}J(k)=\sum_{i=1}^N\frac{\Gamma(p+1)}{\Gamma(p-\beta+1)}\lvert e_{i,k}\lvert^{p-\beta},
				\end{aligned}
			\end{array}
		\end{split}
	\end{equation}
	where $p\geq \beta$, and the second part in the right-hand side of \eqref{008} can be derived as 
\begin{equation}
	\begin{split}
		\begin{array}{rcl}
			\begin{aligned}
				\label{010}
				\Bigg[\frac{\partial\big(\lvert e_{i,k}\lvert\big)}{\partial \hat{\bm h}_k}\Bigg]^{\beta}=\Bigg[\frac{\partial\big(\lvert e_{i,k}\lvert\big)}{\partial e_{i,k}}.\frac{\partial\big( e_{i,k}\big)}{\partial \hat{\bm h}_k}\Bigg]^{\beta}=\big[\text{sgn}(e_{i,k}).(-\bm{x}_{i,k})\big]^{\beta}.
			\end{aligned}
		\end{array}
	\end{split}
\end{equation}

The last term in \eqref{010} has a probability of generating complex values. Therefore, we replace this term with diagonal matrix $-\text{sgn}(e_{i,k})\bm{X}_{i,k}^{\beta-1}\bm{x}_{i,k}$ \cite{cui2025enhanced}, where $\bm{X}_{i,k}$ will be framed as
\begin{equation}
	\begin{split}
		\begin{array}{rcl}
			\begin{aligned}
				\label{011}
				\bm{X}_{i,k} = \text{diag}\big\{\lvert x_{i,k}\lvert, \lvert x_{i,k-1}\lvert,..., \lvert x_{i,k-L+1}\lvert\big\}.
			\end{aligned}
		\end{array}
	\end{split}
\end{equation}

Substituting \eqref{008}, \eqref{009}, and \eqref{010} into \eqref{005}, we obtain
\begin{equation}
	\begin{split}
		\begin{array}{rcl}
			\begin{aligned}
				\label{012}
				\hat{\bm h}_{k+1}=\hat{\bm h}_k+\mu \sum_{i=1}^N g(e_{i,k})\bm{X}_{i,k}^{\beta-1}\bm{x}_{i,k},
			\end{aligned}
		\end{array}
	\end{split}
\end{equation}
where $g(e_{i,k})=\text{sgn}(e_{i,k})\lvert e_{i,k}\lvert^{p-\beta}$, and the term $\Gamma(p+1)/\Gamma(p-\beta+1)$ is absorbed into the step-size $\mu$.

To ensure that \eqref{012} can converge more precisely to its optimal solution when the values of $\bm{x}_{i,k}$ fluctuate sharply, we introduce the normalization term in \eqref{012} and obtain the proposed FoNSPN algorithm, i.e., 
\begin{equation}
	\begin{split}
		\begin{array}{rcl}
			\begin{aligned}
				\label{013}
				\hat{\bm h}_{k+1}=\hat{\bm h}_k+\mu \sum_{i=1}^N \frac{g(e_{i,k})\bm{X}_{i,k}^{\beta-1}\bm{x}_{i,k}}{\lvert\lvert \bm{x}_{i,k}\lvert\lvert_p^p}.
			\end{aligned}
		\end{array}
	\end{split}
\end{equation}
Significantly, when $\beta=1$, the proposed FoNSPN algorithm will degenerate into the traditional NSPN algorithm.
\section{Performance Analysis} 
In this section, we investigate the theoretical parameter ranges (step-size $\mu$ and fractional-order $\beta$) required to ensure stable convergence of the FoNSPN algorithm. For simplicity of calculation, some assumptions are made as follows:

{ Assumption 1 (A1)}: The weak correlation between the signals $\{\bm{x}_{i,k}\}_{i=1}^N$ at different subbands \cite{lee2004improving, 10985768}, and $\{\bm{x}_{i,k}\}_{i=1}^N$ is usually assumed to be a wide-sense stationary random process \cite{10083263}.

{ Assumption 2 (A2)}: $\bm{x}_{i,k}$ and $\widetilde{\bm h}_k$ are asymptotically uncorrelated with $g^2(e_{i,k})$ \cite{al2001adaptive}. $\widetilde{\bm h}_k$ is statistically independent of $\bm{x}_{i,k}$ \cite{10985768}.

{ Assumption 3 (A3)}: The subband noise $n_{i,k}$ has zero mean and variance $\sigma_{n_{i,k}}^2=\sigma_{n_{k}}^2\lvert\lvert \bm{f}_i\lvert\lvert_2^2$ \cite{yin2010stochastic, 10985768}, which is statistically independent of other signals \cite{ye2024optimal}.
\subsection{Bound on the Fractional-Order $\beta$}
By introducing $\bm{E}_k\overset{\bigtriangleup}{=}\hat{\bm h}_{k+1}-\hat{\bm h}_{k}$ and utilizing \eqref{013}, the expectation of its Edclidean norm can be expressed as 
\begin{equation}
	\begin{split}
		\begin{array}{rcl}
			\begin{aligned}
				\label{014} 
				&{\text E}\big\{\lvert\lvert \bm{E}_k\lvert\lvert_2^2\big\}=\mu^2\sum_{i=1\atop {i=j}}^N \text{E}\Big\{\frac{g^2(e_{i,k})\lvert\lvert \bm{X}_{i,k}^{\beta-1}\bm{x}_{i,k}\lvert\lvert_2^2}{\lvert\lvert \bm{x}_{i,k}\lvert\lvert_{p}^{2p}}\Big\} \\&+\mu^2 \sum_{i=1\atop {i\neq j}}^N \text{E}\Big\{\frac{g(e_{i,k})\bm{x}_{i,k}^{\text T}\bm{X}_{i,k}^{\beta-1}}{\lvert\lvert \bm{x}_{i,k}\lvert\lvert_p^p} \frac{g(e_{j,k})\bm{X}_{j,k}^{\beta-1}\bm{x}_{j,k}}{\lvert\lvert \bm{x}_{j,k}\lvert\lvert_p^p}\Big\} .
			\end{aligned}
		\end{array}
	\end{split}
\end{equation}
Here, the final part on the right-hand side of \eqref{014} can be neglected according to A1. Based on A2, \eqref{014} can be approximated as
\begin{equation}
	\begin{split}
		\begin{array}{rcl}
			\begin{aligned}
				\label{015} 
				\text{E}\big\{\lvert\lvert \bm{E}_k\lvert\lvert_2^2\big\}\approx\mu^2\sum_{i=1\atop {i=j}}^N \text{E}\{g^2(e_{i,k})\}\text{E}\Big\{\frac{\lvert\lvert \bm{X}_{i,k}^{\beta-1}\bm{x}_{i,k}\lvert\lvert_2^2}{\lvert\lvert \bm{x}_{i,k}\lvert\lvert_p^{2p}}\Big\}.
			\end{aligned}
		\end{array}
	\end{split}
\end{equation}

Since, the influence exerted of the subband input signal can be regarded as stable and limited \cite{luo2021fractional}. Therefore, the influence exerted by $\text{E}\big\{\lvert\lvert \bm{X}_{i,k}^{\beta-1}\bm{x}_{i,k}\lvert\lvert_2^2/\lvert\lvert \bm{x}_{j,k}\lvert\lvert_p^{2p}\big\}$ on $\text{E}\big\{\lvert\lvert \bm{E}_k\lvert\lvert_2^2\big\}$ is limited. In other words, the value of $\text{E}\big\{\lvert\lvert \bm{E}_k\lvert\lvert_2^2\big\}$ mainly depends on $\mu^2\sum_{i=1}^N\text{E}\{g^2(e_{i,k})\}$. Thus, when FoNSPN algorithm converges to steady-state stage, $\text{E}\big\{\lvert\lvert \bm{E}_k\lvert\lvert_2^2\big\} \rightarrow 0$ is equivalent to $\mu^2\sum_{i=1}^N\text{E}\{g^2(e_{i,k})\} \rightarrow 0$. Based on this, we obtain 
\begin{equation}
	\begin{split}
		\begin{array}{rcl}
			\begin{aligned}
				\label{016} 
				\text{E}\{g^2(e_{i,k})\}=\text{E}\{\lvert e_{i,k}\lvert^{2(p-\beta)}\}.
			\end{aligned}
		\end{array}
	\end{split}
\end{equation}
Since $0\leq p<\alpha$, the stable condition of FoNSPN is
\begin{equation}
	\begin{split}
		\begin{array}{rcl}
			\begin{aligned}
				\label{017} 
				0\leq 2(p-\beta)<\alpha,
			\end{aligned}
		\end{array}
	\end{split}
\end{equation}
which is equivalent to
\begin{equation}
	\begin{split}
		\begin{array}{rcl}
			\begin{aligned}
				\label{018} 
				p-\frac{\alpha}{2}<\beta\leq p.
			\end{aligned}
		\end{array}
	\end{split}
\end{equation}

\textbf{Remark 2}. For the MPE criterion-based FoNSPN algorithm, we usually set $p=\alpha-0.05$ in the $\alpha$-stable noise scenario \cite{ye2024p}. Clearly, $\alpha$ is unknown in practical applications, which can be estimated using our previous research \cite{ye2024p},  Kurtosis’ method \cite{krupinski2006approximated}, and error samples on-line method \cite{zhu2011adaptive}.  
\subsection{Bound on the Step-Size $\mu$}
To determine the theoretical step-size range that guarantees the convergence of the FoNSPN algorithm, we subtract $\bm{h}_0$ from both sides of \eqref{013} and obtain
\begin{equation}
	\begin{split}
		\small
		\begin{array}{rcl}
			\begin{aligned}
				\label{019}
				\widetilde{\bm h}_{k+1}=\widetilde{\bm h}_k-\mu \sum_{i=1}^N \frac{g(e_{i,k})\bm{X}_{i,k}^{\beta-1}\bm{x}_{i,k}}{\lvert\lvert \bm{x}_{i,k}\lvert\lvert_p^p}.
			\end{aligned}
		\end{array}
	\end{split}
\end{equation}

Then, by left-multiplying both sides of \eqref{019} by their respective transposes and taking the expectation, we get
\begin{equation}
	\begin{split}
		\small
		\begin{array}{rcl}
			\begin{aligned}
				\label{020}
				\text{E}\big\{\lvert\lvert\widetilde{\bm h}_{k+1}\lvert\lvert_2^2\big\}&=\text{E}\big\{\lvert\lvert\widetilde{\bm h}_{k}\lvert\lvert_2^2\big\}-2\mu\sum_{i=1}^N\underbrace{\text{E}\Big\{\frac{g(e_{i,k})\widetilde{\bm h}_{k}^{\text T}\bm{X}_{i,k}^{\beta-1}\bm{x}_{i,k}}{\lvert\lvert \bm{x}_{i,k}\lvert\lvert_p^p}\Big\}}_{(a)}\\&+\mu^2\sum_{i=1}^N \text{E}\{g^2(e_{i,k})\} \text{E}\Big\{\frac{\lvert\lvert\bm{X}_{i,k}^{\beta-1}\bm{x}_{i,k}\lvert\lvert_2^2}{\lvert\lvert \bm{x}_{i,k}\lvert\lvert_p^{2p}}\Big\},
			\end{aligned}
		\end{array}
	\end{split}
\end{equation}
where $\text{E}\big\{g^2(e_{i,k})\big\}$=$\text{E}\big\{\lvert e_{i,k}\lvert^{2(p-\beta)}\big\}$, and $g(e_{i,k})\widetilde{\bm h}_{k}^{\text T}\bm{X}_{i,k}^{\beta-1}\bm{x}_{i,k}$ can be equivalently expressed as
\begin{equation}
	\begin{split}
		\begin{array}{rcl}
			\begin{aligned}
				\label{021}
				g(e_{i,k})\widetilde{\bm h}_{k}^{\text T}\bm{X}_{i,k}^{\beta-1}\bm{x}_{i,k}=\frac{g(e_{i,k})e_{i,k}}{e_{i,k}^2}\widetilde{\bm h}_{k}^{\text T}\bm{X}_{i,k}^{\beta-1}\bm{x}_{i,k}e_{i,k}.
			\end{aligned}
		\end{array}
	\end{split}
\end{equation}

Based on \eqref{002}, \eqref{021}, A2, and A3, the term ($a$) in \eqref{020} can be calculated by
\begin{equation}
	\begin{split}
		\begin{array}{rcl}			
			\begin{aligned}
				\label{022}
				&\text{E}\Big\{\frac{g(e_{i,k})\widetilde{\bm h}_{k}^{\text T}\bm{X}_{i,k}^{\beta-1}\bm{x}_{i,k}}{\lvert\lvert \bm{x}_{i,k}\lvert\lvert_p^p}\Big\}=\text{E}\Big\{\frac{\frac{g(e_{i,k})e_{i,k}}{e_{i,k}^2}\lvert\lvert \widetilde{\bm h}_k\lvert\lvert_{\Phi}^2}{\lvert\lvert \bm{x}_{i,k}\lvert\lvert_p^p}\Big\}\\&=\text{E}\Big\{\frac{g(e_{i,k})e_{i,k}}{e_{i,k}^2}\Big\}\text{E}\Big\{\frac{\lvert\lvert \widetilde{\bm h}_k\lvert\lvert_{\Phi}^2}{\lvert\lvert \bm{x}_{i,k}\lvert\lvert_p^p}\Big\},
			\end{aligned}
		\end{array}
	\end{split}
\end{equation}
where $g(e_{i,k})e_{i,k}$=$\lvert e_{i,k}\lvert^{p-\beta+1}$ and $\Phi$=$\bm{x}_{i,k}\bm{X}_{i,k}^{\beta-1}\bm{x}_{i,k}^{\text T}$.

Defining the mean square deviation $\text{MSD}_k\overset{\bigtriangleup}{=}\text{E}\big\{\lvert\lvert\widetilde{\bm h}_{k}\lvert\lvert_2^2\big\}$, and then by inserting \eqref{022} into \eqref{020}, we get
\begin{equation}
	\begin{split}
		\footnotesize
		\begin{array}{rcl}
			\begin{aligned}
				\label{023}
				&\text{MSD}_{k+1}=\text{MSD}_{k}-\\&\sum_{i=1}^N\underbrace{\Big(2\mu\text{E}\Big\{\frac{\lvert e_{i,k}\lvert^{p-\beta+1}}{e_{i,k}^2}\Big\}\text{E}\Big\{\frac{\lvert\lvert \widetilde{\bm h}_k\lvert\lvert_{\Phi}^2}{\lvert\lvert \bm{x}_{i,k}\lvert\lvert_p^p}\Big\}-\mu^2 \text{E}\big\{\lvert e_{i,k}\lvert^{2(p-\beta)}\big\} \text{E}\Big\{\frac{\lvert\lvert\bm{X}_{i,k}^{\beta-1}\bm{x}_{i,k}\lvert\lvert_2^2}{\lvert\lvert \bm{x}_{i,k}\lvert\lvert_p^{2p}}\Big\}\Big)}_{(b)}.
			\end{aligned}
		\end{array}
	\end{split}
\end{equation}
Based on A1, the terms $\text{E}\{\lvert\lvert \widetilde{\bm h}_k\lvert\lvert_{\Phi}^2\}$ and $\text{E}\{\lvert\lvert\bm{X}_{i,k}^{\beta-1}\bm{x}_{i,k}\lvert\lvert_2^2\}$ in \eqref{023} can be calculated by
\begin{equation}
	\begin{split}
		\begin{array}{rcl}
			\begin{aligned}
				\label{025}
				\text{E}\{\lvert\lvert \widetilde{\bm h}_k\lvert\lvert_{\Phi}^2\}=\text{E}\{\lvert {x}_{i,k}\lvert^{\beta+1}\}\text{MSD}_k,
			\end{aligned}
		\end{array}
	\end{split}
\end{equation}
\begin{equation}
	\begin{split}
		\begin{array}{rcl}
			\begin{aligned}
				\label{026}
				\text{E}\{\lvert\lvert\bm{X}_{i,k}^{\beta-1}\bm{x}_{i,k}\lvert\lvert_2^2\}=L\text{E}\{\lvert {x}_{i,k}\lvert^{2\beta}\}.
			\end{aligned}
		\end{array}
	\end{split}
\end{equation}

Clearly, to ensure the FoNSPN algorithm can stable convergence under the mean-square condition, it is necessary that the term $(b)$ in \eqref{023} must be greater than zero. Setting $p-\beta=1$ and substituting \eqref{025} and \eqref{026}, we obtain
\begin{equation}
	\begin{split}
		\begin{array}{rcl}
			\begin{aligned}
				\label{024}
					0<\mu<\text{min}\Big\{\frac{2\text{E}\{\lvert {x}_{i,k}\lvert^{\beta+1}\}\text{MSD}_k}{\big(\sigma_{x_{i,k}}^2\text{MSD}_k+\sigma_{n_{i,k}}^2\big) \frac{L\text{E}\{\lvert {x}_{i,k}\lvert^{2\beta}\}}{\text{E}\{\lvert\lvert \bm{x}_{i,k}\lvert\lvert_p^{p}\}}},i=1,...,N\Big\}.
			\end{aligned}
		\end{array}
	\end{split}
\end{equation}

Unfortunately, \eqref{024} is unable to calculate the upper bound of the step-size owing to depending on iterations. The step-size remains unchanged throughout the adaptive process; therefore, we give a confirmable step-size range.
\begin{equation}
	\begin{split}
		\begin{array}{rcl}
			\begin{aligned}
				\label{024_1}
					0<\mu<\text{min}\Big\{\frac{2\text{E}\{\lvert {x}_{i,k}\lvert^{\beta+1}\}\lvert\lvert \bm{h}_0\lvert\lvert_2^2}{\big(\sigma_{x_{i,k}}^2\lvert\lvert \bm{h}_0\lvert\lvert_2^2+\sigma_{n_{i,k}}^2\big) \frac{L\text{E}\{\lvert {x}_{i,k}\lvert^{2\beta}\}}{\text{E}\{\lvert\lvert \bm{x}_{i,k}\lvert\lvert_p^{p}\}}}\Big\},
			\end{aligned}
		\end{array}
	\end{split}
\end{equation}
where $\text{MSD}_0=\lvert\lvert \bm{h}_0\lvert\lvert_2^2$ is utilized to assess the step-size upper bound. In the case when $\lvert\lvert \bm{h}_0\lvert\lvert_2^2$ approximately equals 1 \cite{ye2024p}, we can draw the conclusion that $\lvert\lvert \bm{h}_0\lvert\lvert_2^2$ only has little influence on the step-size upper bound. Therefore, we have
\begin{equation}
	\begin{split}
		\begin{array}{rcl}
			\begin{aligned}
				\label{024_2}
					0<\mu<\text{min}\Big\{\frac{2\text{E}\{\lvert {x}_{i,k}\lvert^{\beta+1}\}}{\big(\sigma_{x_{i,k}}^2+\sigma_{n_{i,k}}^2\big) \frac{L\text{E}\{\lvert {x}_{i,k}\lvert^{2\beta}\}}{\text{E}\{\lvert\lvert \bm{x}_{i,k}\lvert\lvert_p^{p}\}}},i=1,...,N\Big\}.
			\end{aligned}
		\end{array}
	\end{split}
\end{equation}
\subsection{Steady-State Performance Analysis}
When the FoNSPN algorithm converges to the steady-state stage, i.e., $k\rightarrow \infty$, we have $\text{MSD}_{k+1}=\text{MSD}_{k}$. Setting $p-\beta=1$ and substituting \eqref{025} and \eqref{026} into \eqref{023}, we obtain
\begin{equation}
	\begin{split}
		\begin{array}{rcl}
			\begin{aligned}
				\label{027}
				\text{MSD}_{\infty}=\frac{\mu L\sum_{i=1}^N\frac{\text{E}\big\{\lvert {x}_{i,k}\lvert^{2\beta}\big\}}{\text{E}\big\{\lvert\lvert \bm{x}_{i,k}\lvert\lvert_p^{2p}\big\}}\sigma_{n_{i,k}}^2}{2\sum_{i=1}^N\frac{\text{E}\big\{\lvert{x}_{i,k}\lvert^{1+\beta}\big\}}{\text{E}\big\{\lvert\lvert \bm{x}_{i,k}\lvert\lvert_p^{p}\big\}}-\mu L\sum_{i=1}^N\frac{\text{E}\big\{\lvert {x}_{i,k}\lvert^{2\beta}\big\}}{\text{E}\big\{\lvert\lvert \bm{x}_{i,k}\lvert\lvert_p^{2p}\big\}}\sigma_{x_{i,k}}^2},
			\end{aligned}
		\end{array}
	\end{split}
\end{equation}
where $\sigma_{x_{i,k}}^2$ denotes the variance of the subband input signal $\bm{x}_{i,k}$.

\section{Simulation Result}
\begin{figure*}[htp]
	\centering 
	\includegraphics[scale=0.29] {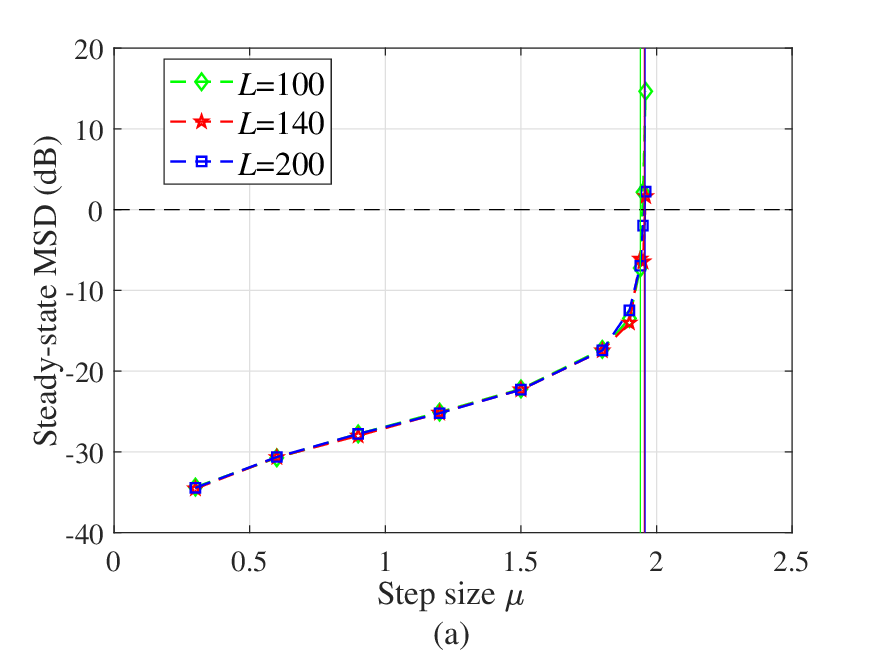}  
	\hspace{0.001ex}	 
	\includegraphics[scale=0.29] {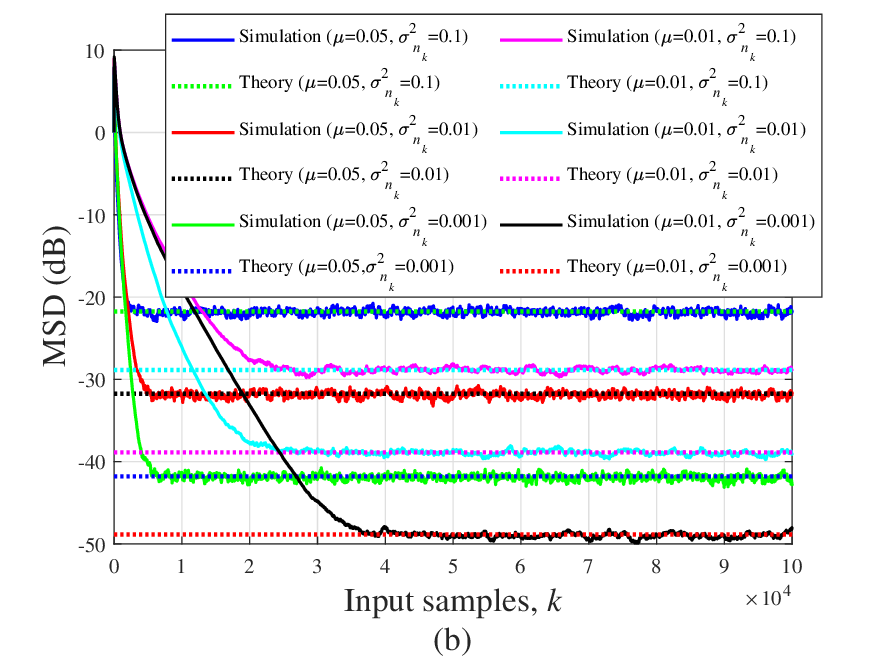}  
	\hspace{0.001ex}									 
	\includegraphics[scale=0.29] {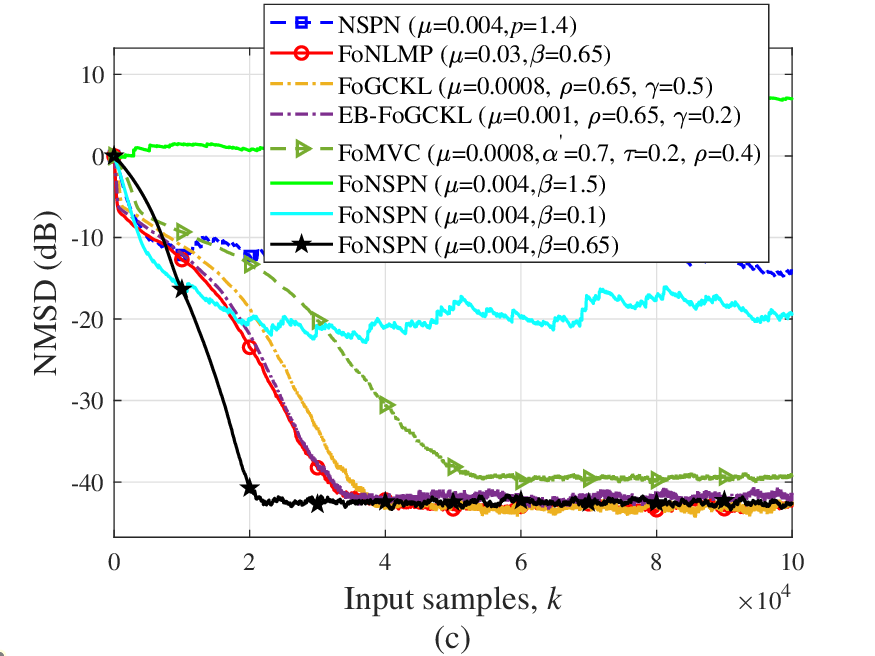}	
	\hspace{0.001ex}									 
	\includegraphics[scale=0.29] {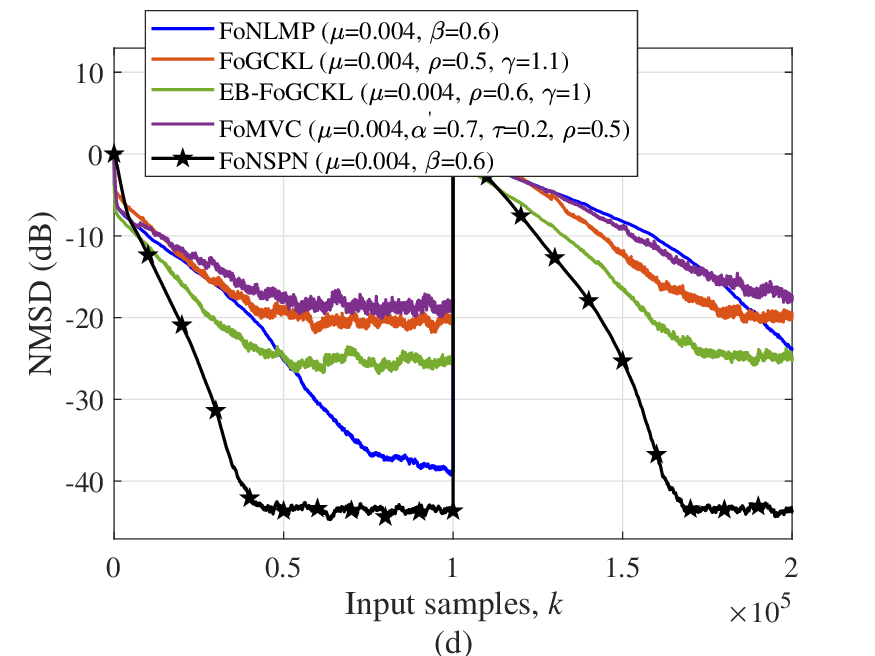}								  
	\caption{Performance verification of the algorithms. (a) Steady-state MSDs versus the step-size when $\sigma_{n_k}^2=0.001$; (b) Theoretical and simulated MSDs versus; (c) $p=0.7$, $\gamma$ denotes the scaling factor, $\rho$ is the order of the fractional derivative, $\tau$ is the versoria parameter, and $\alpha^{'}$ is the fractional-order error parameter; (d) $p=0.7$.  }
	\label{Fig7}
\end{figure*} 
Simulations are performed in this section. Based on the cosine-modulated principle, the analysis filters $\{\bm{f}_i\}_{i=1}^N$ with subband number $N=4$ and length $D=32$ are designed based on the prototype filter. The unknown tap weight vector $\bm{h}_0$ is chosen randomly with length $L=20$. The results are averaged over 50 Monte Carlo trials. In addition, we employ the normalized MSD (NMSD) to evaluate the performance of the algorithms \cite{ye2024optimal}.
\subsection{Gaussian Input}
The input vector $\bm{x}_k$ is generated by a zero-mean Gaussian noise through a first-order auto-regressive model with a pole at 0.999. In Fig. 1 (a), $\bm{h}_0$ is normalized to $\lvert\lvert \bm{h}_0 \lvert\lvert_2=1$. In Figs. 1 (a) and (b), the additive noise $n_k$ is modeled by the Gaussian noise, the expectations in \eqref{024_2} and \eqref{027} are estimated by the ensemble average, and we set $p=2$ and $\beta=1$. In Fig. 1 (a), the steady-state MSDs are calculated by taking the mean of the last ten thousand instantaneous MSD readings and averaging them, clearly, the theoretical upper bound of the step-size closely matches the simulation results. As shown in Fig. 1 (b), the steady-state model \eqref{027} can accurately estimate the steady-state MSD of the FoNSPN algorithm.

The additive noise $n_k$ is modeled by the $\alpha$-stable noise with $\alpha=0.75$ and $\zeta=1/60$ in Figs. (c) and (d). Fig. 1 (c) plots the NMSD curves of the NSPN \cite{ye2024p}, FoNLMP \cite{luo2021fractional}, FoGCKL \cite{cui2025enhanced}, EB-FoGCKL \cite{cui2025enhanced}, FoMVC \cite{abdelrhman2024fractional}, and proposed FoNSPN algorithms. By the experimental parameters and \eqref{018}, the fractional-order $\beta$ that can ensure the proposed FoNSPN algorithm achieves good performance should be restricted to $(0.325,0.7]$. Therefore, when $\beta = 0.1$ and $\beta = 1.5$, the performance of the FoNSPN algorithm declines. Clearly, through proper parameter selection, the FoNSPN algorithm with $\beta=0.65$ obtains a faster convergence than the FoNLMP, FoMVC, FoGCKL, and EB-FoGCKL algorithms.  
\subsection{Cauchy Input}
In Fig. 1 (d), the input signal is generated by a Cauchy noise with $\alpha=1$ and $\zeta=1/10$ through a first-order auto-regressive model with a pole at 0.999. As depicted, the proposed FoNSPN algorithm has better performance than its competing algorithms in terms of convergence rate and steady-state error. In addition, we also verify the tracking capability of the algorithm when the impulse response of the unknown system in the middle of the input sample changes from $\bm{h}_0$ to $-\bm{h}_0$. The results clearly demonstrate that the FoNSPN algorithm exhibits good tracking performance. 
\section{Conclusion}
This paper presents a novel robust subband adaptive filtering algorithm through the integration of FoSGD with the mean $p$-power error (MPE) criterion. The proposed FoNSPN algorithm effectively addresses performance limitations of the conventional NSPN method in environments characterized by $\alpha$-stable noise ($0<\alpha\leq 1$) and complex impulsive noise input. We further establish stability conditions for step-size and fractional-order $\beta$ within the FoNSPN framework. Extensive simulations demonstrate that the proposed algorithm outperforms existing fractional-order adaptive filtering techniques. 

\newpage
\enlargethispage{-5.4cm}
\bibliographystyle{./IEEEtran}
\bibliography{./IEEEabrv,./IEEEexample}
\end{document}